\documentstyle[pra,aps,epsf]{revtex}
\newcommand{\newc}{\newcommand}
\newc{\beq}    {\begin{equation}}
\newc{\eeq}    {\end{equation}}
\newc{\beqa}    {\begin{eqnarray}}
\newc{\eeqa}    {\end{eqnarray}}

\begin{document}

\draft
\twocolumn[\hsize\textwidth\columnwidth\hsize\csname
@twocolumnfalse\endcsname

%\preprint{ cond-mat/000000}
\title{ Quantum Shift Register }
%\author{ Jae-Weon Lee$^1$ , Jae Gil Kim$^2$, and Eok Kyun Lee$^1$\cite{sbnam}  \\}
\author{ Jae-weon Lee\cite{leejw}  and Eok Kyun Lee  \\}
\address{
 Department of Chemistry,  School of Molecular Science (BK 21),
 \\Korea Advanced   Institute of Science and Technology,  Taejon
 305-701, Korea.}

\author{ Jaewan Kim and Soonchil Lee}
\address{
Department of Physics, Korea Advanced Institute of Science and
Technology, Taejon 305-701, Korea}

%\date{\today}
\maketitle

%==================abstract ======================
\begin{abstract}
We consider a  quantum circuit in which  shift and rotation
operations on qubits are performed by  swap gates and
  controlled swap gates. These operations can be useful
 for quantum computers
 performing elementary arithmetic operations such as multiplication
 and a bit-wise comparison  of  qubits.

\end{abstract}

%==========================

%\hskip 3.8cm
%y\vskip2pc
]

%\section {Introduction}

During the last decade there  has been a growing interest in the
theory of quantum computation\cite{summary}. Due to the
  quantum parallelism, quantum computers have the
potential to outperform the classical computers. The shift
operation in classical computers is useful  for a bitwise
manipulation   of data, pseudo-random number
generation\cite{rand}, and  elementary arithmetic operation such
as multiplication and division. So it has been
used as  one of the basic
operations performed in the  CPUs of  classical
computers\cite{computer}.

In this paper, we   investigate the
  function of the quantum shift register
 made of swap gates. The quantum shift register
  means  a  quantum
circuit which can shift
 every data qubit   to the nearest
qubit in a specific direction. We further study its
applications to arithmetic calculation
and bit-wise operations on two quantum registers.
  As is well known,
 the swap gate consists of three CNOT gates\cite{cnot} and
can be realized by
NMR\cite{NMR}, which is useful for reordering of
qubits such as in the quantum Fourier transform\cite{qft}.

 Fig.1 shows a classical shift
left (shift to the direction of higher bit ) register using D
flip-flop circuit. Each flip-flop contains a classical bit and
moves this  to the next higher flip-flop simultaneously at a
specific clock signal such as a rising edge.

\begin{figure}[Fig1]
\epsfysize=6cm \epsfbox{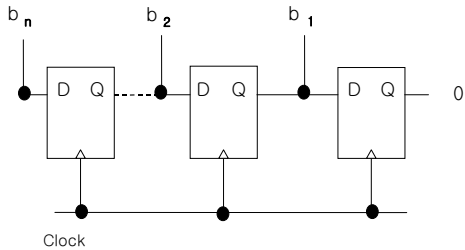} \caption[Fig1]{\label{fig1label}
Schematic diagram showing
 the classical n-bits shift-left register composed of D flip-flop.

}
\end{figure}

The input(Q) of each flip-flop is connected
to the output (D) of
the next bit in lower  flip-flop in this classical shift
 register.
The lowest bit is usually set to 0.
Due to the irreversibility of the
 classical shift  operation, the information originally
stored in the highest bit($b_n$)will be lost after the shift.
Each flip-flop is a feedback circuit
which is constituted  of classical logic gates such as
XOR gates.
Unfortunately, since the feedback required for the flip-flop  is
not a unitary operation, it is impossible to make a
quantum shift register by  straightforward  conversion
of the
classical logic gates in the flip-flop  into corresponding quantum logic
gates.

 Let us consider the quantum  shift register consisted of  swap
 gates instead of the flip-flops. In Fig.2,
 a quantum
 circuit which can perform both
 shift-left and rotation-left on  a $n$-qubit data
  is presented.

\begin{figure}[Fig2]
\epsfysize=6cm \epsfbox{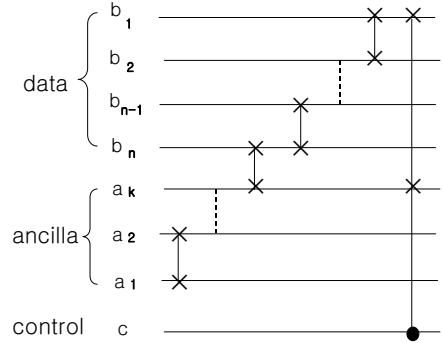} \caption[Fig1]{\label{fig1} The
quantum shift  and rotation left register which is composed
of the swap gates and the controlled swap
(Fredkin)  gate and $k$ ancilla qubits  operating $k$ shifts on
n-qubits .

}
\end{figure}

 It can be easily shown
  that sequential exchanges of qubits
 is equivalent to  the shift operation.
  Since a unitary
evolution is reversible, the information contained in the quantum
shift register does not  disappear contrary to the
case of the
classical shift register\cite{delete}.
Hence, to preserve the
information, we need  at least $k$ extra qubits
to operate  $k$ shifts.
If we denote the ancilla qubits as $|a\rangle =|a_1
a_2...a_k\rangle $ and the data qubits as $|b\rangle=| b_1 b_2
...b_{n}\rangle $, then the  initial
 state of the shift register(omitting
the control qubit) is $|a\rangle\otimes |b\rangle \equiv |a_1 a_2
...a_k; b_1 b_2 ...b_{n}\rangle $.
In this quantum circuit, the types of the operation
activated is  governed by the control qubit.
 To activate  the shift operation,
 the control qubit $c$ is set
to $|0\rangle$ ,  and to activate the rotation operation,
 $c$ is set
to $|1\rangle$.

The  ancilla
qubits can be initially set to  $|0\rangle$ for convenience.
 In this case, each  swap  operation transforms
 the state of the  register as
$|a_1 a_2...a_k; b_1 b_2 ...b_{n}\rangle   \rightarrow
|a_2a_1...a_k;b_1 b_2 ...b_{n}\rangle \dots \rightarrow |
a_2...a_k b_n; b_1 b_2... a_1 b_{n-1}\rangle  \dots \rightarrow |
a_2...a_k b_n; a_1 b_1 b_2 ...b_{n-1} \rangle $.
 Since the control qubit $c$ is set
to $|0\rangle$ ,  the last swap between $a_k$ and $b_1$ is
inhibited.
After one shift(i.e.,after $n+k-1$ swaps),
 the ancilla qubits
becomes $|a_2...a_k b_n \rangle $.
 Note that, during the swaps and
the shift, the ancilla qubits remain disentangled from the data
qubits.

 Second, let us
consider the rotation left operation.
In this case,
the control bit is set to $|1\rangle$ to allow the swap between the last
ancilla qubit and the first data qubit after the shift-left
operation on the data qubits. Then we obtain $|a_1 a_2...a_k; b_1
b_2 ...b_{n}\rangle \rightarrow \dots
 | a_2...a_k b_n; a_1 b_1 b_2 ...b_{n-1}
\rangle  \rightarrow | a_2...a_k a_1; b_n b_1 b_2 ...b_{n-1}
\rangle $, which is equal to output
obtained by operating
left-rotation   both on   $n-$data qubits and
on  $k-$ancilla qubits.

%Finally, if the control qubit is set to a
%superposition of
%  $|0\rangle$ and
%$|1\rangle$, then the final output is
%a superposition of the state made of
% the shift state  and rotation  on the
% initial state.

 It is straightforward to generalize the shift register
 to  a  shift-right
and a rotation-right register by simply reversing the order of the
swaps.

Here, we present some applications of the shift register.
We use
a conventional abbreviation which omits  explicit
representation of the ancilla qubits.
 With the quantum shift register a quantum computer can easily
perform
 multiplication\cite{mult}.
Let us consider  the multiplication of
$n$ numbers($a_i$) in a state
$A=(n)^{-1/2}\sum_{i=0}^{n-1}|a_i\rangle$ by a classical binary
number $1100_{(2)}$. This is achieved  by
performing addition of two outputs obtained by
operating shift-left two times on $A$ and
three times  on $A$ provided that
the shift register has enough size to
 accommodate the results.
 The reason is as follows.
Generally, if a multiplier  $l$ is denoted by
$l=l_{k-1}\cdots l_{0(2)}=\sum^{k-1}_{p=0} 2^p l_p$ in a binary form,
 then multiplication of $a_i$ by $l$ can be written as
 \beq
 a_i\times l=\sum_{p=0}^{k-1} a_i 2^p {l_p}.
 \label{times}
 \eeq
 Here the multiplication  of $a_i$ by $2^p$
 can be performed by operating
   shift-left $p$-times
 on $n$ numbers $(a_i,i=0\cdots n-1$).
 Therefore, summation of all those terms obtained by
performing shift-left  operation   $p$-times
 on $a_i$ when
$l_p=1$  results in the
multiplication of $a_i$ and $l$.

\begin{figure}[Fig3]
\epsfysize=6cm \epsfbox{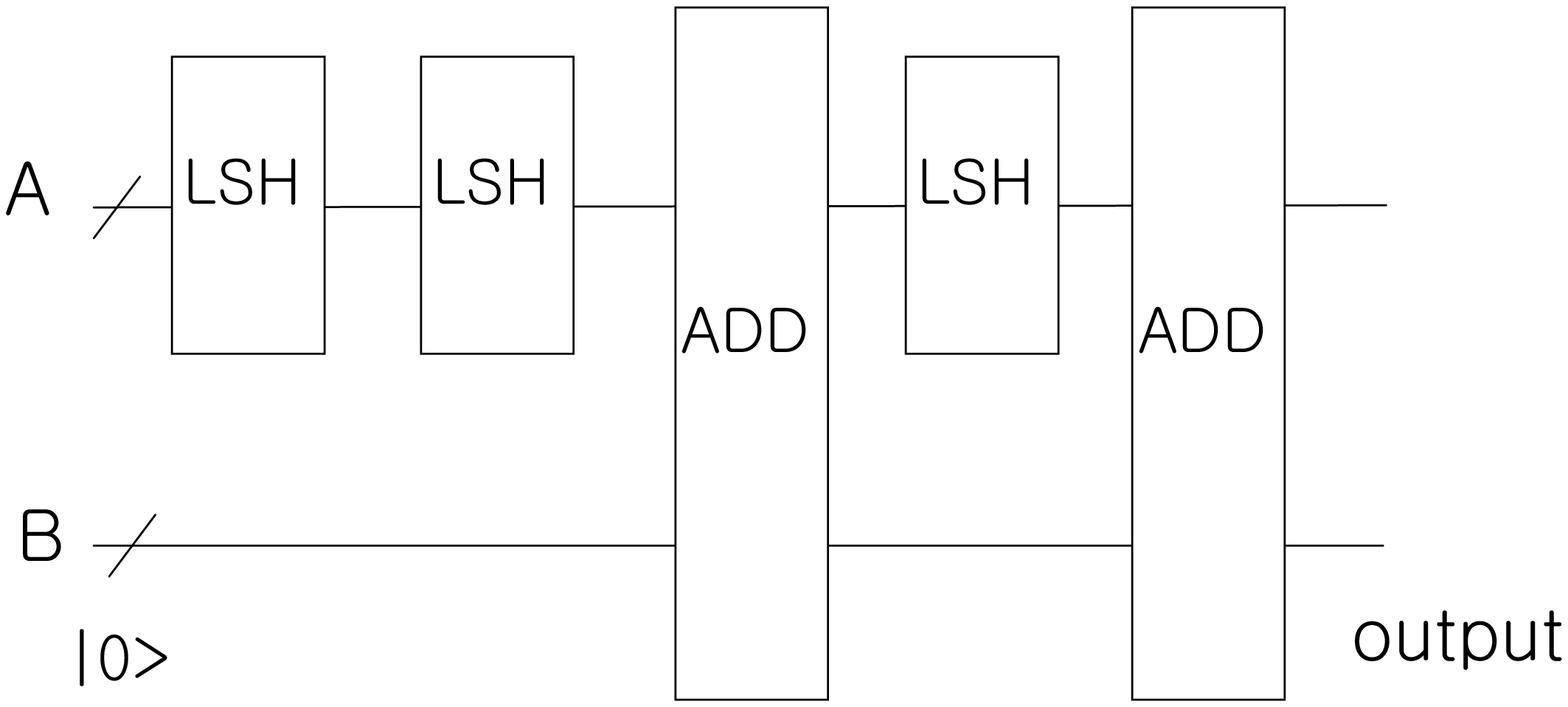} \caption[Fig3]{\label{fig3}
The scheme of quantum algorithm performing multiplication
$a_i$ by a number
$1100_{(2)}$. Here LSH is the quantum left shift register and ADD
is a quantum adder. The  lines with a slash denote qubit wires.

}
\end{figure}

However, due to the No-Cloning theorem
we can not make a
copy of $A$ necessary for the
generation of $2^2 A$ and $2^3A$\cite{noclon}.
This difficulty can be overcome
by preparing a work state $B=|0\rangle$,
defining
 a quantum adder $ADD(A,B)=(A, A+B)$\cite{mult}, and
adding $B$  to the output
obtained by  performing  appropriate shift operation on
 $A$   as
shown in Fig.3.

By using  the quantum adder we can make
$B$ state contain the desired final
results. In this respect, $B$
plays the role of an `accumulator' in a
classical CPU. Following illustration
offers a clear explanation.
For the initial state $A=(|0\rangle_A +|1\rangle_A)/\sqrt 2$
, $B=|0\rangle_B$
and  the same multiplier $l=1100_{(2)}$,  one
can  get entangled states $(|0000\rangle_A|0000\rangle_B +
|1000\rangle_A|1100\rangle_B)/\sqrt 2 $ (omitting ancilla qubits)
after three shifts and two additions on the initial
$A$ and $B$, respectively.
 It is easy to verify that
the number of  operations required
for multiplication  of
$A=(2^n)^{-1/2}\sum_{m=0}^{2^n-1}|m\rangle$ by a $k$ digits
positive  binary number using the quantum shift register is
$O(k)$ ( each shift is composed of $n+k-1$ swaps. See
Fig.2).  On the other hand,  classical computers require
$O(k)$ shifts and additions for every $2^n$ numbers($m$), separately.
Thus the total number of operations  required for this  multiplication
in a  classical computers
is $O(2^n k)$.
Hence, a quantum computer will outperform
a classical computer in multiplications,
 if the quantum shift register
  can be realized.
Of course, with the quantum computer, we can obtain superposition
of all results of multiplication  but only one result is
available  after
measurement, contrary to the classical computers.
However,  the
multiplication circuit can be used as
a part of larger quantum algorithm  which
does not require measurement just after the multiplications, just
as the modular multiplication is a part of  the Shor's factoring
algorithm\cite{shor,mult}. So embedded in a larger
quantum algorithm,  the multiplication circuit
using the quantum shift register can show exponential speed up
compared to the classical multiplication circuit.

Now, it is straightforward  to extend
our arguments to the case where the $k$ digits binary
multipliers themselves are $m$ numbers ($c_j, j=1,2,...,m$) contained in a
superposition of states
$C=(m)^{-1/2}\sum_{j=1}^{m}|c_j\rangle$.(See Fig.4)

\begin{figure}[Fig4]
\epsfysize=6cm \epsfbox{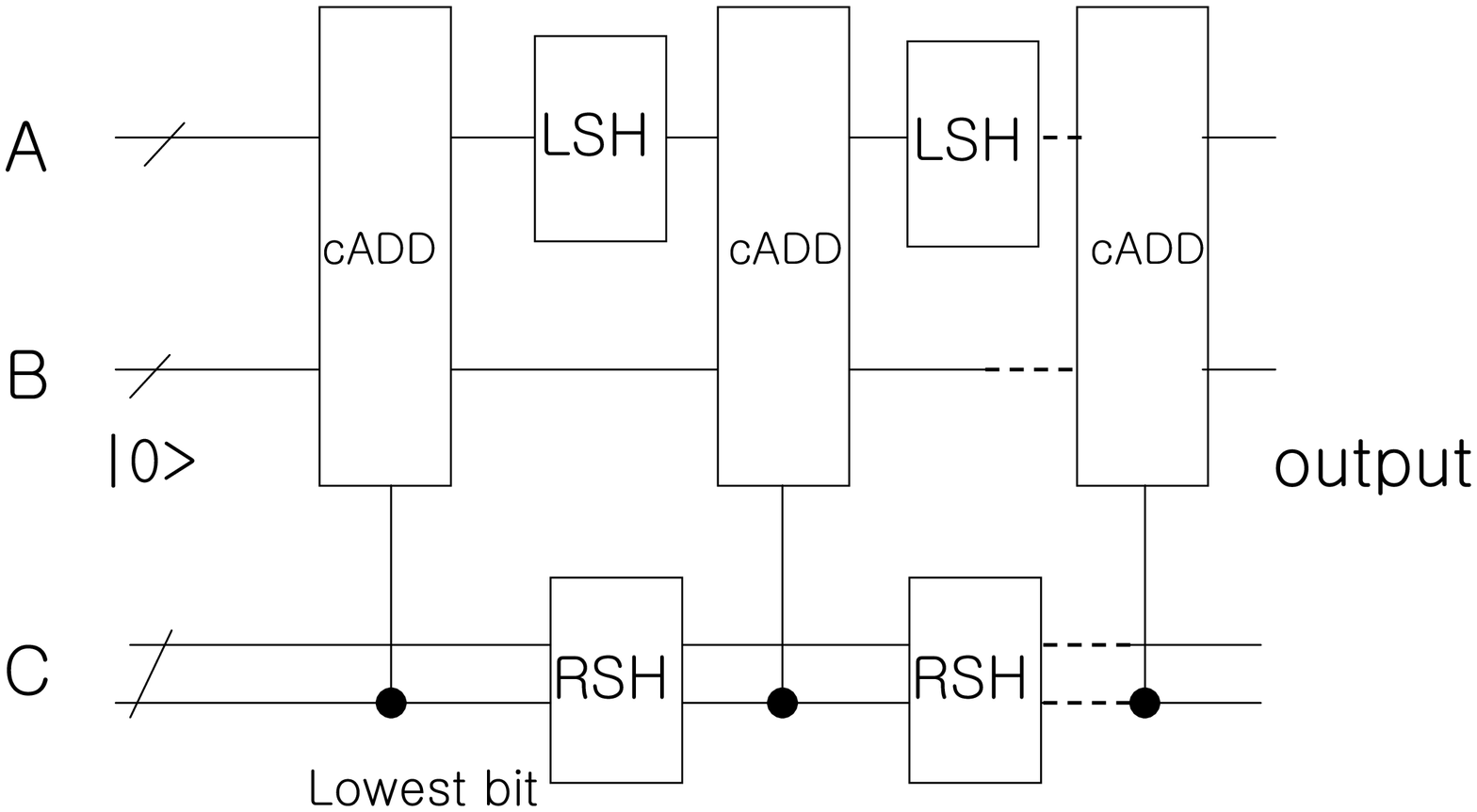} \caption[Fig4]{\label{fig4}
Quantum algorithm performing multiplication of two states A and
C. Here RSH is the quantum right shift register and cADD is a
controlled quantum adder. The  lines with a slash denote qubit
wires. The lowest qubit of $C$ is shown separately for clarity.
}
\end{figure}

Let us consider the operations depicted in Fig. 4.
To begin with, performing a conditional quantum addition of
$A=(n)^{-1/2}\sum^{n-1}_{i=0} |a_i\rangle$ and $B=|0\rangle$
only when the lowest qubit of $C$ is $|1\rangle$
leads to
$B=(nm)^{-1/2}\sum_{j=1}^{m}\sum^{n-1}_{i=0}
|a_i c^0_j\rangle$. Here
$c_j$ is
denoted in a binary form $c_j^{k-1}\cdots c_{j(2)}^{0}$.
Then
% following the arguments below the eq.(\ref{times}),
one can find that
 repetitions of
 above conditional quantum addition of $A$ and $B$
 only when the lowest qubit of $C$ is $|1\rangle$
after operating  shift-left operation on $A$ and
shift-right on $C$
  is equivalent to
  performing  the
quantum parallel multiplications of all numbers $a_i$ and $c_j$
simultaneously and storing the result in $B$.

Fig.4   also
shows another possible application of the shift register. It can
be used to select out a specific qubit in the register by
shifting the qubit , for example, to the lowest qubit as the
right-shift register does.

As  is shown in Fig.1,
 the shift register in our model  requires many swaps. Hence, it will be
interesting to consider the quantum shift register performing the
shift operation simultaneously on each qubits to speed up the
calculation using, for example, flying qubits\cite{flying}.

In summary, we have constructed
the  quantum shift and rotation register
utilizing  swap gates and a controlled swap gate with
the help of the ancilla qubits. The quantum circuit
 can be used for  fast numerical operations such as
multiplication and qubit selections.

%\subsection*{}
\vskip 1cm
%We thank KRISS members for their warm hospitalities at KRISS.
%Specially JWL thanks Dr. Y. H. Lee for his kindness and SBN thanks
%Drs J. C. Park and Y. K. Park for various discussions.
This work
is supported by BK21.

\end{document}